\documentclass[manuscript=article]{achemso}

\mciteErrorOnUnknownfalse % a fix to compile with MikTeX (Neset)

\usepackage{ifthen}
\usepackage[usenames]{color}
\usepackage{amssymb,mathbbol,amsmath,amsbsy}
\usepackage{amsfonts}
\usepackage{dsfont}
\usepackage{amsmath}
\usepackage{graphicx}
\usepackage[utf8]{inputenc}
\usepackage{float}

\usepackage{lipsum}  
\usepackage{xcolor}
\usepackage{soul}
\usepackage{multirow}
\usepackage{lscape}

\newcommand{\Trace}{\mathrm{Tr}}
\newcommand{\trans}{\mathcal{T}}
\newcommand{\GG}{\mathcal{G}}
\newcommand{\HH}{H}
\newcommand{\II}{I}
\newcommand{\lmfp}{\ell_\mathrm{mfp}}
\newcommand{\lloc}{\ell_\mathrm{loc}}
\newcommand{\lint}{\ell_\mathrm{inter}}
\newcommand{\Nlayer}{N_\mathrm{layer}}
\newcommand{\Tmulti}{\mathcal{T}_\mathrm{multi}}
\newcommand{\Tmono}{\mathcal{T}_\mathrm{mono}}
\newcommand{\rhomulti}{\rho_\mathrm{multi}}
\newcommand{\rhomono}{\rho_\mathrm{mono}}

\newcommand{\haldun}[1]{{{#1}}}
\newcommand{\ciz}[1]{{\color{red}}}
\DeclareUnicodeCharacter{0301}{\'{e}}

\author{Mustafa~Ne\c{s}et~\c{C}{\i}nar}
\affiliation{%
Department of Materials Science and Engineering, Izmir Institute of Technology, 35430 Urla, Izmir, Turkey.
}
\author{Aleandro~Antidormi}
\affiliation{%
	Catalan Institute of Nanoscience and Nanotechnology, CSIC and The Barcelona Institute of Science and Technology, Campus UAB, Bellaterra, 08193 Barcelona (Cerdanyola del Vallès), Spain.
}
\author{Viet-Hung~Nguyen}
\affiliation{%
	Institute of Condensed Matter and Nanosciences, Universit\'e catholique de Louvain (UCLouvain), B-1348 Louvain-la-Neuve, Belgium.
}
\author{Alessandro~Kovtun}
\affiliation{%
	Consiglio Nazionale delle Ricerche, Istituto per la Sintesi Organica e la Fotoreattività, (CNR-ISOF), via Gobetti 101, 40129 Bologna, Italy.
}
\author{Samuel~Lara~Avila}
\affiliation{%
	Department of Microtechnology and Nanoscience, Chalmers University of Technology, Kemivägen 9, 41296 Gothenburg, Sweden.
}
\author{Andrea~Liscio}
\affiliation{%
	Consiglio Nazionale delle Ricerche, Istituto per la microelettronica e microsistemi, Roma Unit (CNR-IMM), via del fosso del cavaliere 100, 00133 Roma, Italy.
}
\author{Jean-Christophe~Charlier}
\affiliation{%
	Institute of Condensed Matter and Nanosciences, Universit\'e catholique de Louvain (UCLouvain), B-1348 Louvain-la-Neuve, Belgium.
}
\author{Stephan~Roche}
\email{stephan.roche@icn2.cat}
\affiliation{%
	Catalan Institute of Nanoscience and Nanotechnology, CSIC and The Barcelona Institute of Science and Technology, Campus UAB, Bellaterra, 08193 Barcelona (Cerdanyola del Vallès), Spain.
}
\altaffiliation{%
ICREA–Institució Catalana de Recerca i Estudis Avançats, 08010 Barcelona, Spain
}

\author{H\^{a}ldun~Sevin\c{c}li}
\email{haldunsevincli@iyte.edu.tr}
\affiliation{%
	Department of Materials Science and Engineering, Izmir Institute of Technology, 35430 Urla, Izmir, Turkey.
}

\newcommand{\baslik}{Towards Optimized Charge Transport in Multilayer Reduced Graphene Oxides}
\title{\baslik}

\begin{document}

\begin{abstract}
In the context of graphene-based composite applications, a complete understanding of charge conduction in multilayer reduced graphene oxides (rGO) is highly desirable. However, these rGO compounds are characterized by multiple and different sources of disorder depending on the chemical method used for their synthesis. Most importantly the \haldun{precise} role of interlayer interaction in promoting or jeopardizing electronic flow remains unclear. Here, thanks to the development of a multiscale computational approach combining first-principles calculations with large scale transport simulations, the transport scaling laws in multilayer rGO are unraveled, explaining why diffusion worsens with increasing film thickness. In contrast, contacted films are found to exhibit an opposite trend when the mean free path \ciz{is} \haldun{becomes} shorter than the channel length, since conduction becomes predominantly driven by interlayer hopping. These \ciz{theoretical outcomes} \haldun{predictions} are favourably compared with experimental data and open a road towards the optimization of graphene-based composites with improved electrical conduction. 
\end{abstract}

\maketitle

Understanding charge transport in multi-layered van der Waals materials has become an attractive and challenging problem, in the perspective of both fundamental and applied research on graphene-based composites~\cite{ferrari_nanoscale_2015,mohan_compositesb_2018}.
Indeed, graphene-related materials (including chemically disordered graphene like rGO) have shown remarkable capability to improve charge and thermal conductivities of many insulating flexible materials such as organic polymers, suggesting them as the privileged filler material to reinforce, diversify and improve the properties and performances of traditional materials used in wearables, flexible electronics, conducting textiles and thermoplastics. However, the accurate understanding of the microscopic mechanisms leading to transport in these complex systems is still \ciz{a debated issue} \haldun{a matter of debate}. Indeed, recent experimental studies~\cite{turchanin_acsnano_2011,silverstein_nanoscale_2019,kovtun_acsnano_2021} suggested the key role played by various types of defects, as well as interlayer interaction in the transport mechanisms taking place inside the graphene-based composites.

More specifically, transport characteristics in multilayered rGO~\cite{kovtun_acsnano_2021} were observed to \ciz{move} \haldun{change} from a conventional Efros-Shkloskii variable range hopping (ES-VRH) to a temperature-dependent power-law regime, \ciz{when the number of layers increase in the stack} \haldun{with increasing the number of stacked layers}. Such findings do not apparently depend on any length scale of the system, since they are observed in both micrometric networks of few nanosheets partially overlapping, as well as in centimeter‐scale thin films built from billions of rGO nanosheets randomly stacked.
\ciz{Besides, the resulting localization length is also found to increase (by three orders of magnitude) with both the aromatic content and the thickness of the multilayered system as well, while being completely independent on the lateral size of the layer in the stack.}
\haldun{Besides, the resulting localization length is also found to increase (by three orders of magnitude) with both the aromatic content and the thickness of the thin films as well, while being roughly independent on the lateral size nanosheet. %Thus, in the first approximation, we can safely neglect the role of the rGO nanosheet edges describing networks of rGO nanosheets randomly stacked as multilayered structures.
Accordingly, the main contribution to transport properties in multilayered rGO with random stacking likely stems from a bulk contribution, with marginal nanosheet edge effects.~\cite{kovtun_acsnano_2021}}

\begin{figure*}[t]
	\centering
	\includegraphics[width=1.0\textwidth]{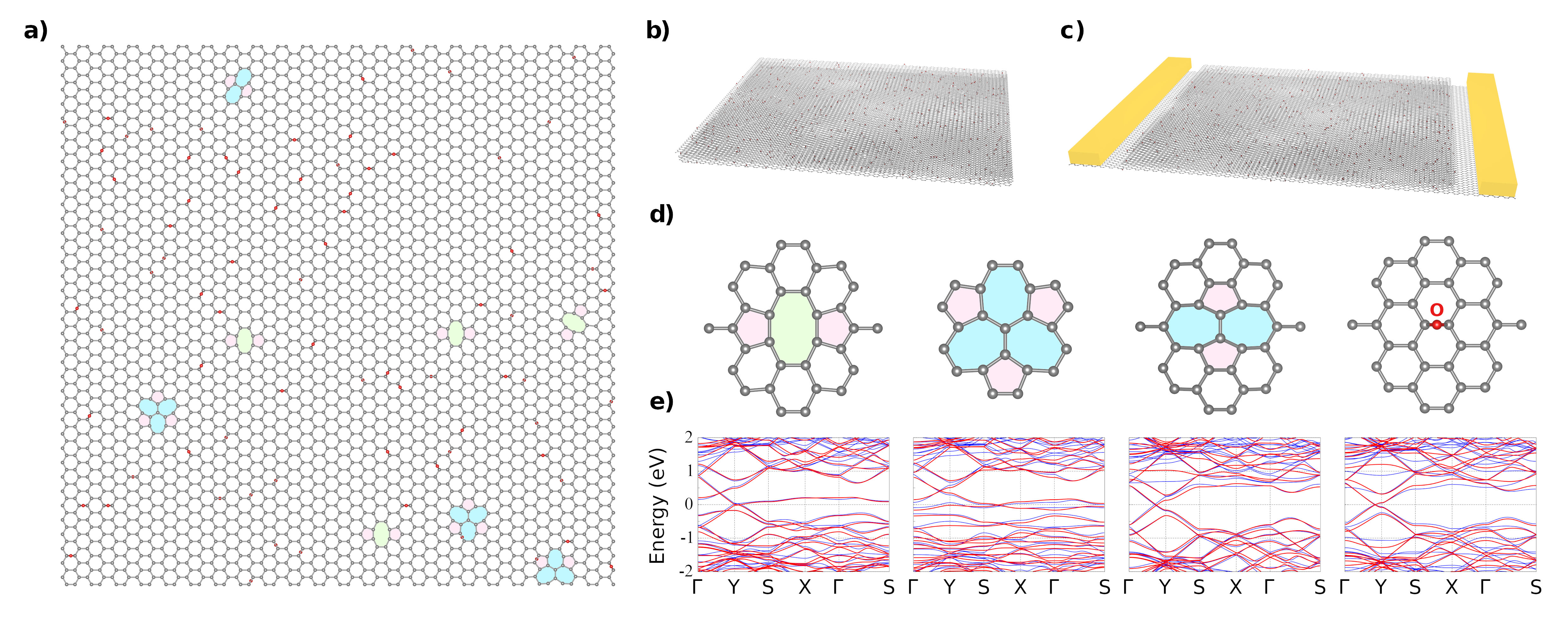}
	\caption{
		Reduced graphene oxide hosts different types of defects that are randomly distributed (a), \haldun{where carbon atoms are colored with gray, oxygen with red. Pentagon, heptagon and octagons are indicated with red, blue and green shades.} Charge transport calculations in typical multilayered rGO models are investigated in rGO using (b) the Kubo formalism to estimate the bulk electronic conductivity, and (c) the Landauer-B\"uttiker approach in a conventional device geometry. 
		(d) Atomic structures of various incorporated defect types (pristine divacancy (585), reconstructed divacancy (555-777), Stone-Wales (55-77) and epoxy defects) and their corresponding tight-binding models (e) extracted from {\it ab initio} band-structure calculations.
	}
	\label{fig1}
\end{figure*}

\ciz{In the present work, charge transport in rGO thin films is investigated theoretically}
\haldun{Here, we investigate charge transport in rGO thin films}
using state-of-the-art modelling techniques and \ciz{the corresponding results are compared}
\haldun{compared the results} 
with experimental measurements. 
The multilayered rGO models are constructed using a rectangular ribbon-like geometry with a width of 20 nm, and periodic boundary conditions in the transverse direction. Disorder is introduced \ciz{through} \haldun{by} incorporating random distributions of chemical defects (Fig.~\ref{fig1}a) such as divacancies (0.29\%), Stone-Wales defects (0.01\%) and epoxides (4.7\%). Such chemical nature and densities of defects have been extracted from atomistic samples of \ciz{reduced graphene oxide} \haldun{rGO} obtained by \haldun{classical molecular simulations of} \ciz{simulating} the thermal reduction process of graphene oxide (GO) sheets.~\cite{Antidormi}
\ciz{via classical molecular simulations}
More specifically, 
\ciz{the thermal annealing protocol described in Ref. is employed }
\haldun{to produce the models, we employ the thermal annealing protocol described in Ref.~\citenum{Antidormi},}
since it \ciz{yields to} \haldun{reproduces} the main structural features observed in \ciz{studied} \haldun{measured} rGO samples. In these MD simulations, large-scale GO samples with an initial equivalent concentration of epoxide and hydroxil groups, amounting to a \ciz{C/O} \haldun{O/C} ratio of 35\%, were annealed at \ciz{300~$^\mathrm{o}$C} \haldun{900~$^\mathrm{o}$C}.
\haldun{At the end of the annealing process, the final concentration of oxygen atoms was found to be approximately 5\%, in excellent agreement with the experimental results reported in Ref.~\citenum{kovtun_acsnano_2021}.}

These chemical defects are known to induce lattice distortion and charge redistribution locally around their spatial location.~\cite{Cresti2008,zhang_natphys_2009,wang_nanoscale_2012,ma_prl_2014,gao_apl_2015,zhang_carbon_2016}.
Since these detrimental effects are local, rGO models with defect concentrations similar to those investigated in this study can be modeled using a conventional tight-binding (TB) Hamiltonian (as for graphene) but with specific adjustments made locally around the defect position.
Within such a framework, the Hamiltonian of the rGO system presents a simple form allowing to further perform large scale transport calculations in realistic samples containing more than $10^6$ atoms and with varying defect concentrations. The parametrized $p_z$ TB Hamiltonian reads
\begin{eqnarray}
	H_0=\sum\limits_{\langle p,q\rangle} \gamma_{pq}c_p^\dagger c_q,
\end{eqnarray}
where in-plane couplings are limited to nearest neighboring interactions \ciz{only}. In contrast with other models in the literature (see Refs.~\citenum{leconte_acsnano_2010,lherbier_prb_2012}), the effects of chemical defects are described by adjusting properly the TB parameters to recover {\it ab initio} results. The optimization of these TB parameters for each specific single defect are obtained by fitting first-principles electronic band structures of graphene supercell containing a single defect (see Supporting Information).

Importantly, the change in C-C bond length due to  the lattice distortion around the defects is accounted by computing the hopping energies as a function of C-C bond length $(r_{pq})$ and defining $\gamma_{pq}=\gamma_0\exp[-\beta(r_{pq}/r_0-1)]$ with \haldun{nearest-neighbor hopping energy} $\gamma_0=-2.6$~eV, \haldun{nearest-neighbor distance} $r_0 = 1.42$~$\textup{\AA}$ and \haldun{the decay parameter} $\beta = 3.37$.~\cite{pereira_prb_2009} 
The local doping due to the localized states induced by defects and impurities is also included by modulating onsite energies as a distance-decay function centered at the defect positions. 
Since the atomic positions are altered around defects, the parametrization for interlayer coupling also requires to account for changes in interatomic distances compared to Bernal graphite (AB-stacking).
Consequently to determine TB couplings in presence of varying bond lengths around structural defects, an exponential decay-based model formula is used following $\gamma(r) = \gamma_1 \exp(\beta_z(1-r/z))$ with \haldun{interlayer coupling energy} $\gamma_1$=0.36 eV, \haldun{the corresponding decay parameter} $\beta_z$=24.99 and interlayer distance $z$=3.34~$\textup{\AA}$.~\cite{morel_prb_2010,reich_prb_2002}
Further details 
\haldun{and comparison against density functional theory and other methods are given} in the Supporting Information.
The different types of considered defects used in the model are illustrated in Fig.~\ref{fig1}d. 
The electronic bands obtained from TB parametrization are in excellent agreement with first-principles density functional theory results (Fig.~\ref{fig1}e).
\haldun{It is worth mentioning that the concentration of defects is large enough so as to randomize carriers’ momenta (more than a hundred scatterings within interlayer diffusion length) and hence the transport properties are not expected to depend on the relative twist angle between layers.}

%-----------------------------------------------------------------------------
% Transport Section
%-----------------------------------------------------------------------------

\begin{figure}[ht!]
	\centering
	\includegraphics[width=1.0\textwidth]{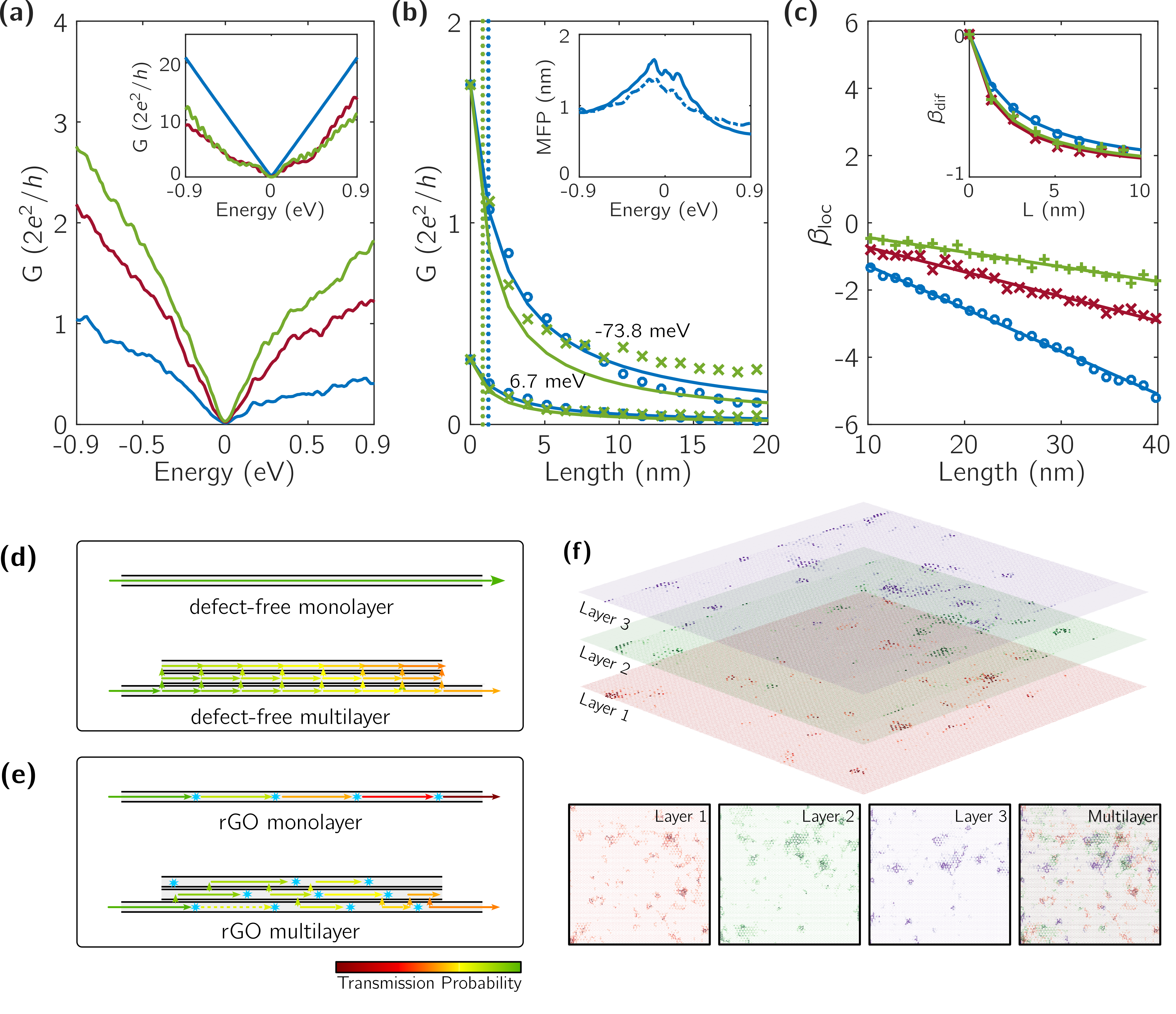}
	\caption{
		(a) Electronic conductance of mono/bi/tri-layer rGO devices (blue, red, green curves, respectively) as a function of energy. The conductance is found to increase with the number of rGO layers, in contrast with stack of pristine graphene (inset).
		(b) Electronic conductance of mono- and tri-layered rGO systems for $E=6.7$~meV (lower region) and $E=-73.8$~meV (upper region) estimated by LB (markers) and Eqn.~\ref{eqn:conductance-mfp} (solid lines). Mean free paths used in Eqn.~\ref{eqn:conductance-mfp} are determined by using KG method.
		Diffusive versus localized regimes in multilayered rGO devices are distinguished by referring to the $\beta$-function (c).
		Transport regimes in pristine and and defective stacks are illustrated in (d-e).
		(f) Local density of states (LDOS) around the Fermi level in multilayer rGO is shown for a relatively smaller sample. LDOS for individual layers and the multilayer are shown in the below panels. Red, green, blue correspond to individual layers from bottom to top. 
	}
	\label{fig2}
\end{figure}

\ciz{The} Kubo-Greenwood (KG) and Landauer-B\"uttiker (LB) transport formalisms are used to study the electronic conductivity and conductance in \ciz{several} multilayered rGO models and for \ciz{different} \haldun{varying} transport geometries (see Fig.~\ref{fig1}b-c, respectively).
LB method allows to include the charge injection from contact electrodes (Fig.~\ref{fig1}c) whereas the KG method gives access to bulk properties~\cite{fan_linear_2020} (Fig.~\ref{fig1}b). These two techniques enable contrasting bulk properties with ``device'' related transport (see Supporting Information for details). Note that neither electron-phonon coupling nor many-body effects are included in the present study.

Fig.~\ref{fig2}a presents the energy-dependent conductance for the mono-, bi- and trilayer rGO for identical defect concentration, calculated within the LB approach.
When layers are free \ciz{of} \haldun{from} defects (pristine graphene), adding new layers will act as supplemental scattering source, hence reducing the total conductance (see Fig.~\ref{fig2}a-inset).
Indeed, in defect-free monolayer graphene, transport is ballistic and a maximum V-shaped conductance curve is obtained. When \ciz{additional} \haldun{more} clean layers are stacked on top of this graphene layer, new scattering channels are opened and an interface resistance is formed.
Consequently, the conductance of bi- and tri-layers (multilayered stack) is reduced below the one of pristine monolayer graphene.
Actually, such conductance decay is not specific to disorder-free systems but \haldun{is} also observed in low-defect concentrations as well (see Fig.~\ref{fig_supp_lowdefect}).
Importantly, this reduction of the electronic transmission is found when the electrodes are attached to the bottom layer and it decreases in a non-monotonic way with the number of added layers.
This behavior can be understood as driven by complicated interference processes taking place across the layers and throughout the central region. 
%\colorbox{red}{REFERENCE MISSING HERE !} 

In sharp contrast, in strongly disordered rGO systems, the electronic behavior is opposite, in the sense that the higher the number of added layers on the stack, the larger the conductance (Fig.~\ref{fig2}a-main frame). This counterintuitive result suggests that the interlayer hopping is opening more conductive channels once the localization length of bottom layer is short enough due to strong in-plane disorder.
The two opposite roles played by interlayer coupling in defect-free and defective multilayers are \ciz{illustrated} \haldun{pictured} in Fig.~\ref{fig2}d-e.

To better understand the electronic transport mechanisms in mono- and multilayered rGO systems, we perform a scaling study of the conductance to better differentiate between diffusive and localized regimes. Figure~\ref{fig2}b shows results for the mono- to multilayered rGO systems presenting the same \ciz{concentration of defects} \haldun{defect density}. Clear evidences for different scaling behaviors is observed (note that conductance values from LB simulations are marked with circles (monolayer) and crosses (trilayer)). In the diffusion regime, the conductance dependence with the length is theoretically expected to scale as function of the mean-free-path ($\lmfp$) and the defect-free conductance ($G_\mathrm{Graphene}$) of the system (at a given energy) following
\begin{eqnarray}
\label{eqn:conductance-mfp}
\frac{1}{G}=\frac{1}{G_\mathrm{Graphene}}
\left(
1+\frac{2}{\pi}\frac{L}\lmfp
\right),
\end{eqnarray}
where $L$ is device length.~\cite{datta_1995}

For short enough channel length the computed conductance scaling is indeed well described by the diffusive formula (Eqn.~\ref{eqn:conductance-mfp}), allowing the extraction of $\lmfp$ from simulations. For the chosen disorder features, $\lmfp$ in the range of 1-2 nm are obtained (Fig.~\ref{fig2}b-inset), regardless the chosen transport formalism. As the number of layers is increased, $\lmfp$ decreases slightly (Fig.~\ref{fig_supp_mfp}).
We note that we here report the scaling behavior of $G$ values for two selected energies (6.7~meV and -73.8~meV) but the trends are similar at all energies around the charge neutrality point (CNP).

\begin{figure}[t]
	\includegraphics[width=0.65\textwidth]{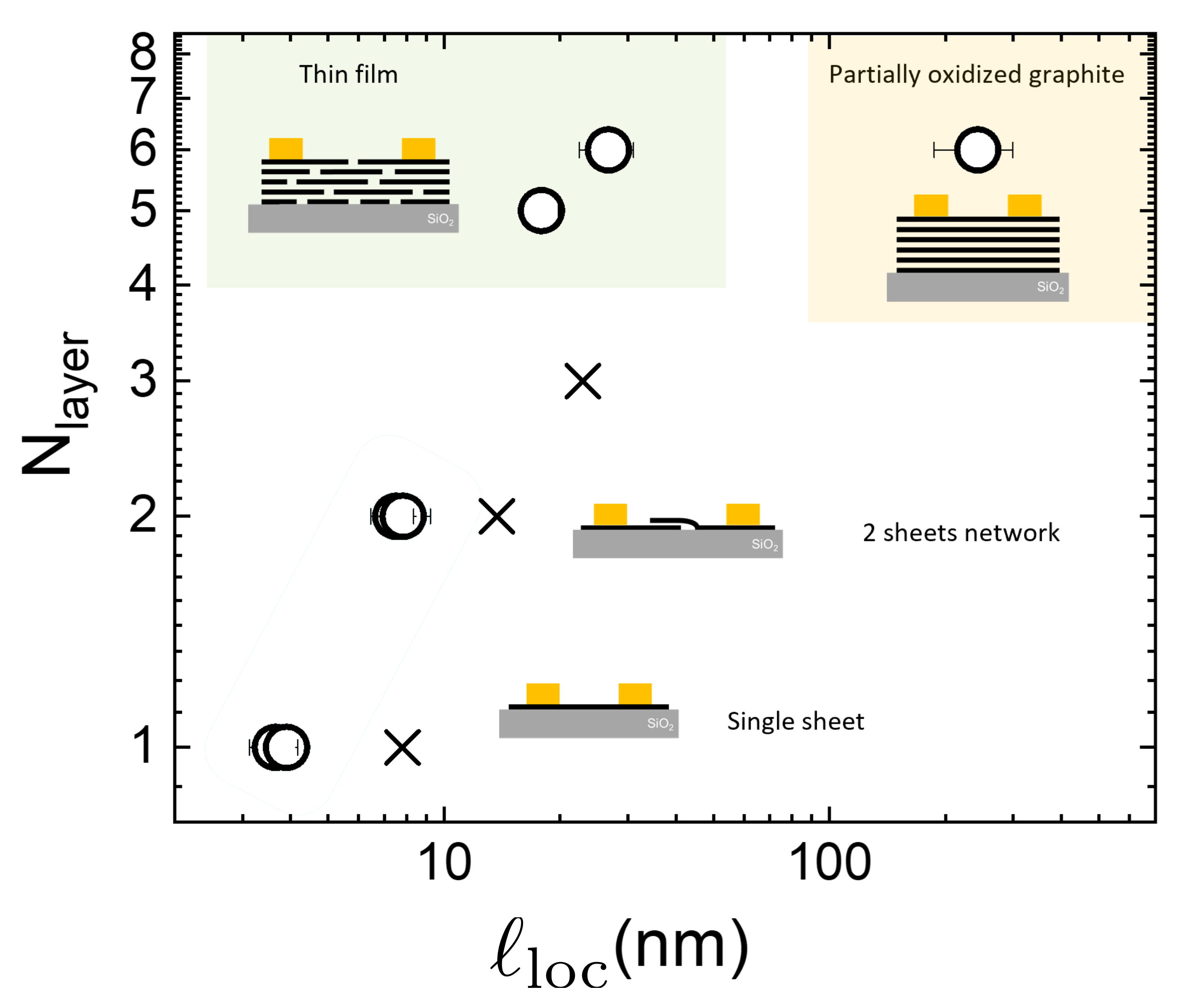}
	\caption{
		\ciz{Experimental localization lengths for varying number of layers. $\Nlayer=6$ stands for partially oxidized graphite and the remaining are randomly stacked rGO sheets.}
		\haldun{
			Localization lengths for varying number of layers.
			Circles represent experimental values, crosses are simulation results.
			$\Nlayer\leq6$ stands for randomly stacked rGO shhets, and the last data point is for partially oxidized graphite. Experimental}
		$\lloc$ values are calculated using different dielectric constants. In the case of single or few stacked layers, we estimate $\epsilon_r= 2.5$ from the average of the dielectric constant of the substrate (SiO$_2$ = 3.9) and vacuum/air (= 1)~\cite{jang_prl_2008}, while for RGO thin film we assume $\epsilon_r = 3.5$,~\cite{jung_jphyschemc_2008} and $\epsilon_r= 15$ for partially oxidized graphite. 
	}
	\label{fig:localization}
\end{figure}

We \ciz{now} \haldun{further} observe that for channel length $L>10$~nm, \ciz{computed} \haldun{the} conductance values for the monolayer are below the diffusion curve (which are plotted by solid lines). This pinpoints the onset of localization effects. However for the trilayer case, a puzzling sudden change of the conductance behavior is seen for a channel length of 10~nm. To further substantiate this striking difference of the conductance scaling between monolayer and multilayer\haldun{ed stack}, we investigate the localization regime and capture quantitative information about the localization lengths. LB simulations are performed on rGO systems with long channel lengths and using the scaling function defined as $\beta={\partial \ln G}/{\partial \ln L}$.~\cite{abrahams_prl_1979}
Substituting the corresponding expressions for diffusion and localization, $\beta$ can be written as $\beta_{dif}=-(1+\pi\lmfp/2L)^{-1}$ in the diffusive regime, whereas $\beta_{loc}{=}-L/\lloc$ in the localization regime with $\lloc$ being the localization length.
Both scaling functions ($\beta_{dif}$ and $\beta_{loc}$) are presented in Fig.~\ref{fig2}c versus the device length.
The solid curves represent predictions of the above-mentioned analytical formulas, whereas the markers are directly obtained from the simulation (for an energy of 6 meV) without referring to $\lmfp$ or $\lloc$, since the scaling functions can be expressed as $\beta_{dif}=G/G_\mathrm{Graphene}-1$ and $\beta_{loc}\propto\ln G$.

On one hand, for the short-channel rGO devices, the diffusive regime is confirmed since the $1/L$ behavior is observed when using $\lmfp$ (see Fig.~\ref{fig2}c-inset). On the other hand, localization regime is obtained for longer-channel rGO devices, and the evaluated localization lengths varies with the number of layers:
7.8~nm for monolayer,  13.7~nm for bilayer and  22.9~nm for trilayer. Thus, when increasing the number of layers, hopping transport gives rise to larger transmission amplitudes compared to diffusion because of the enhancement of $\lloc$, \haldun{in very good agreement with the experimental findings as shown in Fig.~\ref{fig:localization}.}

Such behavior well agrees with experimental findings achieved on rGO devices with different
number of layers ($\Nlayer$) and similar chemical structure (\emph{i.e.} sp$^2$  content $96\pm1\%$).~\cite{kovtun_acsnano_2021}
Figure~\ref{fig:localization} collects the results obtained comparing \ciz{six} \haldun{seven} different systems ranging from the single nanosheet ($\Nlayer = 1$) to a flake of partially oxidized graphite ($\Nlayer = 6$). Differently, all the other devices \ciz{with intermediate thickness} \haldun{($2\leq\Nlayer\leq6$)} are assemblies of rGO sheets randomly stacked. According with the results reported in Ref.~\citenum{kovtun_acsnano_2021}, all of the six devices reveal ES-VRH transport mechanisms at low temperatures (10~K$<$T$<$100~K). Thus, the corresponding localization length ($\lloc$) is calculated using the temperature-dependent electrical resistivity curves
\ciz{$\rho(T)=\rho_0\exp\{A\epsilon_r\lloc T\}^{-1/2}$} 
\haldun{$\rho(T)=\rho_{0,VRH}\exp\{A\epsilon_r\lloc T\}^{-1/2}$ }
where \haldun{$\rho_{0,VRH}$ is a prefactor for resistivity,}
$\epsilon_r$ is the relative dielectric constant of the material and
\ciz{$A=0.021$~$\mu$m$^{-1}$K$^{-1}$ }
\haldun{$A=2.8e^2/4\pi\epsilon_0k_B=0.021$~$\mu\mathrm{m}^{-1}\mathrm{K}^{-1}$}
for a 2D system (see Supporting Information for more details). 
\ciz{Up to $\Nlayer=5$ we find} \haldun{rGO devices have} $\lloc$ values varying from ca 4~nm to \ciz{17~nm} \haldun{30~nm}, showing the same trend and the same order of magnitude when compared with the simulations. It is noteworthy to underline that in the case of partially oxidized graphite the corresponding $\lloc$ value is one order of magnitude larger amounting to 250~nm, clearly evidencing the combined role of the crystalline structure and the dielectric properties. 
Such aspects are out of the scope of this work and some details related to the experimental setup and the device characterizations are reported in the Supporting Information.

The substantial increase in the localization lengths is thus \ciz{an interesting} \haldun{a key} transport feature to distinguish between various multilayered rGO devices and \haldun{which} can be rationalized withing the variable range hopping (VRH) framework.~\cite{mott_philmag_1969}
Indeed, at zero temperature, the energy separation between two localized states should be very small to enable significant hopping, with probability proportional to $e^{-2\alpha R}$ (with $1/\alpha$ being the attenuation length for the localized states and $R$ being the spatial separation of states). 
The hopping probability between two states $\psi_A$ and $\psi_B$, which have similar energies and are localized at the bottom layer well-separated from each other, should thus be strongly enhanced if a third state $\psi_C$ is localized at the upper layer and between $\psi_A$ and $\psi_B$.
Therefore layers with localized states have enhanced transmission when stacked.
In Fig.~\ref{fig2}f, the local density of states (LDOS) is plotted for a trilayer rGO sample. It is clearly visible that states localized at different layers tend to fill the spatial gaps when they are superimposed.
This can qualitatively explain the enhancement of $\lloc$ with number of layers.

\begin{figure}[t]
	\includegraphics[width=0.55\textwidth]{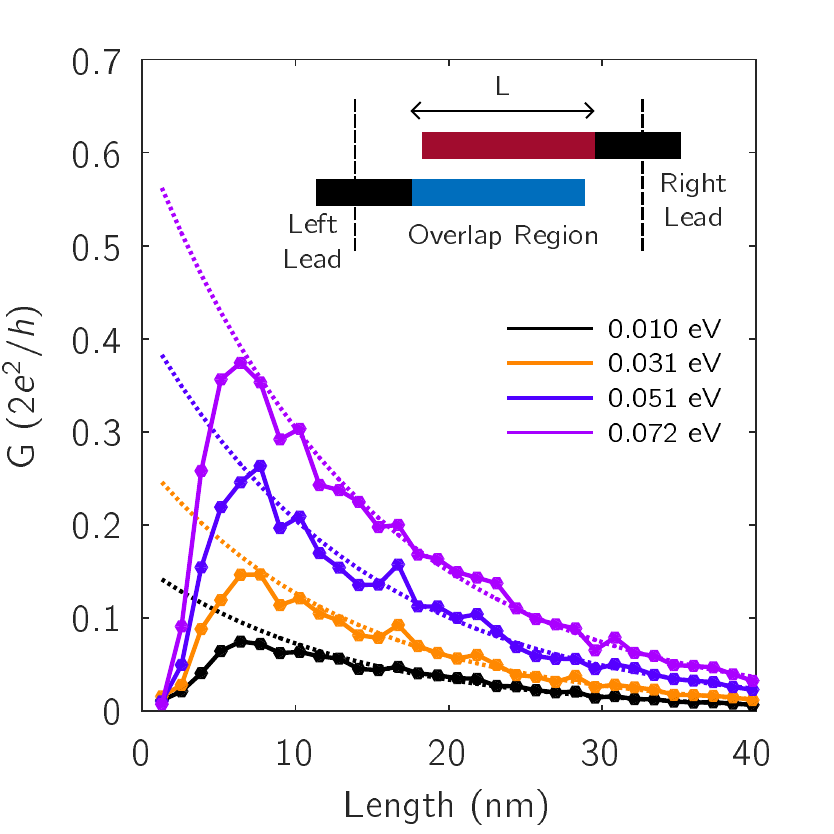}
	\caption{Conductance as a function of overlap length in bilayer reduced graphene oxide when the electrodes are connected to different layers. Dotted curves represent exponential decay.}
	\label{fig:overlap}
\end{figure}

The interlayer coupling thus plays a crucial role in multilayer transport. In low defect concentrations, it is a source of scatterings \ciz{and suppresses} \haldun{which impedes} transport, whereas in highly-defective rGO multilayers, it tends to promote \ciz{longer $\lloc$} \haldun{longer state delocalization}.
The contact geometry is also a major factor in transport across the layers.
To gain further information about the combined roles of interlayer coupling and contact geometry, 
we simulate a situation where the device is made from overlapping bilayer rGO in which the injection and collection takes place at different layers (\ciz{see the} inset of Figure~\ref{fig:overlap}).
We compute the conductance with different overlap distances, $L$, which is also the length of the central region.
As shown in Fig.~\ref{fig:overlap}, the conductance rapidly increases with $L$ at short distances because the overlap area enhances the probability for a carrier to diffuse from one layer to the other.
Differently\haldun{,} for overlap distances longer than 10~nm, the conductance decays exponentially with $L$, indicating that the localization behavior prevails over the interlayer diffusion.
The maximum conductance is achieved at around 7~nm, which marks the interlayer diffusion length ($\lint$).
We have checked the dependence of $\lint$ in clean structures, and found very similar values, which proves that $\lint$ is dictated by the strength of interlayer coupling and not by the disorder content.
We can actually estimate a length scale for interlayer diffusion as $hv_F/\gamma_1=11.5$~nm, where the Fermi velocity is $v_F=10^6$~m/s and the interlayer coupling strength is $\gamma_1=0.36$~eV.
The comparison $\lint\gg\lmfp$ in the simulated structures  is the reason behind the fact that $\lmfp$ is weakly affected from the increase in the number of layers.
On the other hand, since $\lint$ is comparable with the monolayer localization length, there is room for localized carriers \ciz{spread to} \haldun{to further spread over the} neighboring layers.
\ciz{In order to} \haldun{To} clarify the effect of contact geometry on transport, we have considered the electrodes 
\ciz{to have the same number of layers with the central region.}
\haldun{with the same number of layers as in the central region.}
In this geometry, \haldun{as expected,} the conductance increases with the number of layers both in clean and defective samples, \ciz{as expected} but the behavior of neither $\lmfp$ nor $\lloc$ are considerably affected. 

\begin{figure}[t]
	\includegraphics[width=0.5\textwidth]{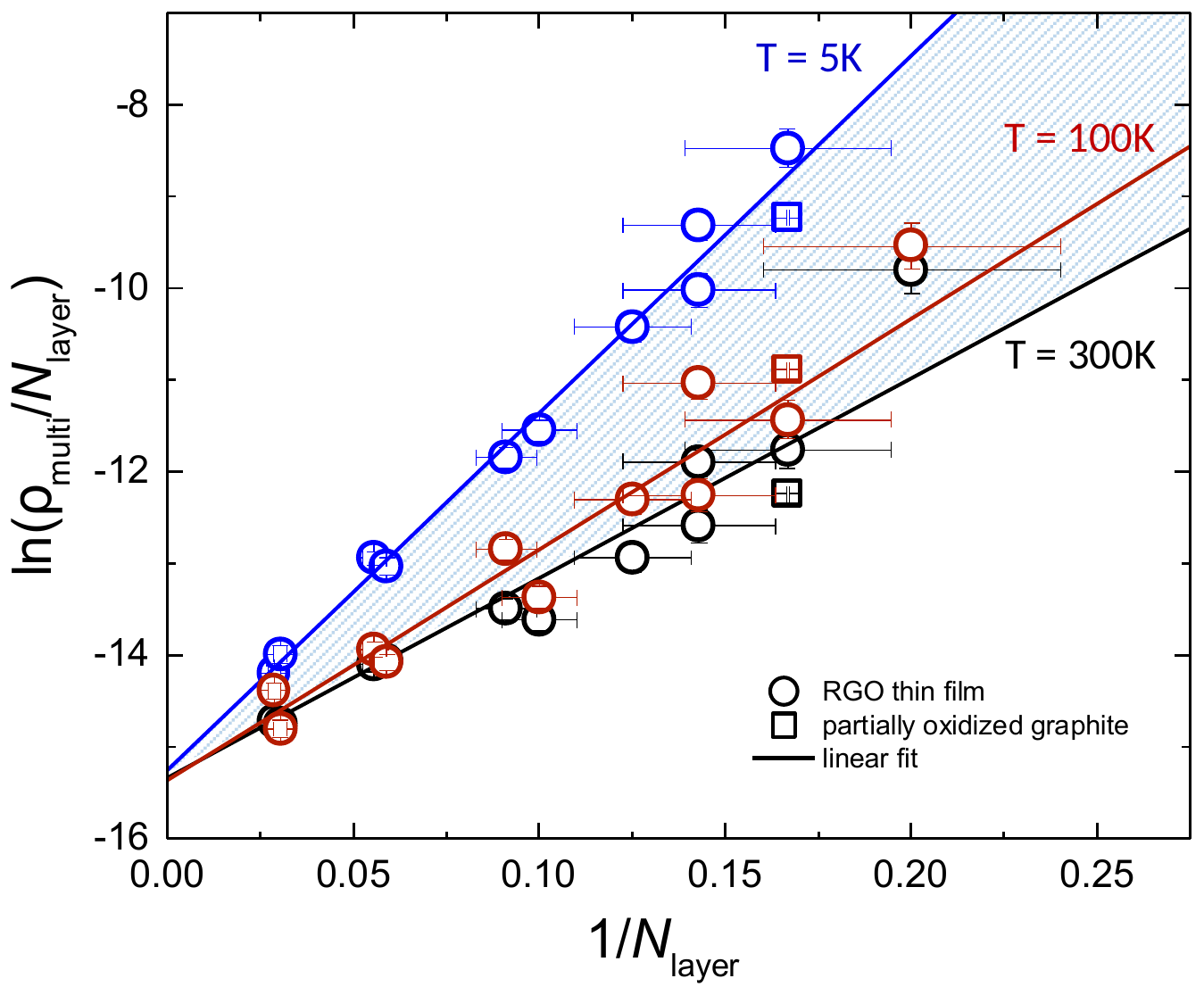}
	\caption{
		Scaling of multilayer resistivity with the number of layers is plotted as
		$\ln(\rho_\mathrm{multi}/\Nlayer)$ versus $1/\Nlayer$. Experimental data of multilayer rGO (circles for devices reported in Ref.~\citenum{kovtun_acsnano_2021} and square for partially oxidized graphite) acquired at different temperatures show linear dependence, in good agreement with the theoretical prediction for scaling (cf. Eqn.~\ref{eqn:multilayerscaling}). All the linear fitting curves calculated at different temperatures are included between the two curves acquired at 5~K and 300~K (dashed area).
	}
	\label{fig:fit}
\end{figure}

Finally, based on these findings, the scaling of the resistivity with the number of layers can be expressed quantitatively. In highly defective rGO multilayers $\lmfp$ changes only slightly with the number of layers but $\lloc$ increases linearly with $\Nlayer$.
Using the fact that $\lloc$ is proportional to the number of transmission channels and  $\lmfp$,\cite{thouless_jphysc_1973,leconte_prb_2011} one can approximate the transmission probability across multilayer rGO in terms of that of the monolayer as $\Tmulti=\Tmono^{1/\Nlayer}$. 
Correspondingly the resistivities  ($\rho$) satisfy the following relation,
\begin{equation}
	\frac{1}{\Nlayer}
	\left(
	\frac{\rhomulti}{\rho_0}
	\right)
	=
	\left(
	\frac{\rhomono}{\rho_0}
	\right)^{1/\Nlayer},
	\label{eqn:multilayerscaling}
\end{equation}
where $\rho_0=hA_\mathrm{mono}/2e^2L$ is a system wide constant $A_\mathrm{mono}$ being the cross section area for a monolayer. Namely, we predict a linear dependence of $\ln(\rhomulti/\Nlayer)\sim 1/\Nlayer$
for film thickness larger than the mean free path. 
For the sake of comparison, we analyze a data set of 11 rGO devices with similar chemical structure (i.e. sp$^2$ content~=~96$\pm1\%$) and film thickness ranging from 2~nm to 13~nm, i.e. $5 < \Nlayer < 35$.
\haldun{Note that, we do not need to assume any values for  $A_\mathrm{mono}$ and $\rho_0$ or measure them,
%in our simulations or in Fig.~\ref{fig:fit}, 
but they are used in order to relate resistivity and conductance so as to compare experimentally measured values with simulated ones through the scaling relation (Eqn.~\ref{eqn:multilayerscaling}). }

Figure~\ref{fig:fit} shows the correlation plot $\ln(\rhomulti/\Nlayer)$ vs $1/\Nlayer$ for each device displaying the resistivity values acquired at different temperatures: 5~K, 100~K and 300~K, the lowest, an intermediate and the highest measured, respectively. A total of datasets corresponding to 43 different temperatures were analyzed and the remaining 40 curves – not depicted in the figure – are included between the two curves acquired at 5~K and 300~K (dashed area). In all the cases we observe a linear trend: $y = m\cdot x+q$ in excellent agreement with Equation~(\ref{eqn:multilayerscaling}). Moreover, we obtain that the slope $m = \ln(\rho_\mathrm{mono}/\rho_0)$ decreases with increasing temperature, while the $y-$intercept ($q$) is a constant value corresponding to $\rho_0 = \exp(q) = (2.1\pm0.2)\times10^{-7}$~$\Omega\cdot$m.
Similarly, a linear behavior is achieved in the case of rGO devices with lower amount of the aromatic content 
%(77\% and 86\%, see Figure \haldun{SX}),
(77\% and 86\%, see Fig,~\ref{fig:supp_correlation})
where the resistivity values increase with the oxidation degree, as expected.
Summarizing the experimental findings, $\rhomono$ is the single-layer resistivity (temperature-dependent) while $\rho_0$ does not depend on the temperature, being therefore a kind of resistivity scaling factor only depending on the aromatic content of the device.

\section{Conclusion}

We have reported quantum simulations on realistic models of multi-layered rGO which reveal the complex interplay between disorder and interlayer interactions in dictating the dominant transport mechanism.
Depending on the concentration of defects, multilayer interaction can enhance or suppress the system conductance, which results from the competition between the mean free path $\lmfp$ and the interlayer diffusion length $\lint$. When about 5\% of the carbon atoms are involved in defected regions,
$\lint$ becomes much longer than $\lmfp$.
In that case, intralayer scattering largely dominates over interlayer diffusion, leading to a weak dependence of $\lmfp$ on $\Nlayer$. 

On the other hand, $\lloc\sim \lint$, so that a localized state in one of the layers has enough extension for tunneling to an adjacent one.
If $\lint$ was much larger than $\lloc$, the tunneling rates would be much smaller.
Once $\lint\sim\lloc$, tunneling rates are appreciable and charge delocalization is promoted.
While $\lmfp$ is weakly dependent on $\Nlayer$, $\lloc$ increases with $\Nlayer$ as expected from a generalization of the Thouless relationish for one-dimensional conductors.
This unprecedented interplay between transport length scales is a specific result of 2D layered nature of the multilayer rGO systems. Such mechanism enables hopping transport to overcome the diffusion limit, which is usually the upper bound in bulk systems.
Our theoretical analysis enables us to derive a novel scaling rule, which is in perfect agreement with experimental data at various temperature, and consistent with the Thouless relationship.
The fundamental findings of this study are not limited to multilayered reduced graphene oxide but could find applications in other two-dimensional stacks as well.

%%%%%%%% acknowledgements
\medskip%%
{\noindent{\textbf{Acknowledgements.}}}
%\par
%\noindent 
The authors acknowledge support from the  Flag-Era JTC  2017  project  `ModElling
Charge and Heat trANsport in 2D-materIals based Composites$-$ MECHANIC'.
MN\c{C} and HS acknowledge support from T\"UB\.ITAK (117F480). AA and SR are supported by - MECHANIC reference number: PCI2018-093120 funded by Ministerio de Ciencia, Innovacion y Universidades and the European Union Horizon
2020 research and innovation programme under Grant Agreement No.
881603 (Graphene Flagship). ICN2 is funded by the CERCA Programme/
Generalitat de Catalunya, and is supported by the Severo Ochoa program
from Spanish MINECO (Grant No. SEV-2017-0706).
V.-H.N. and J.-C.C. acknowledge financial support from the F\'ed\'eration Wallonie-Bruxelles through the ARC on 3D nano-architecturing of 2D crystals (N$^\circ$16/21-077), from the European Union's Horizon 2020 Research Project and Innovation Program --- Graphene Flagship Core3 (N$^\circ$881603), and from the Belgium FNRS through the research project (N$^\circ$T.0051.18).
Computational resources have been provided by the CISM supercomputing facilities of UCLouvain and the CECI consortium funded by F.R.S. -FNRS of Belgium (N$^\circ$2.5020.11).
\haldun{Authors are particularly grateful to Prof.~Paolo Samorì and Marco Gobbi for the scientific assistance to prepare the devices, Valentina Mussi for some supporting measurements and for enlightening discussions.}

%\bibliography{biblio.bib}
\clearpage

\providecommand{\latin}[1]{#1}
\makeatletter
\providecommand{\doi}
  {\begingroup\let\do\@makeother\dospecials
  \catcode`\{=1 \catcode`\}=2 \doi@aux}
\providecommand{\doi@aux}[1]{\endgroup\texttt{#1}}
\makeatother
\providecommand*\mcitethebibliography{\thebibliography}
\csname @ifundefined\endcsname{endmcitethebibliography}
  {\let\endmcitethebibliography\endthebibliography}{}

\newpage
\clearpage
%%%%%%%%%%%%%%%%%%%%%%%%%%%%%%%%%%%
%\clearpage
%\onecolumngrid
\setcounter{equation}{0}
\setcounter{figure}{0}
\setcounter{page}{1}
\renewcommand\theequation{S\arabic{equation}}
\renewcommand\thefigure{S\arabic{figure}}
\renewcommand\thepage{S\arabic{page}}

\newcommand{\yazarlar}{Mustafa~Ne\c{s}et~\c{C}{\i}nar, Aleandro~Antidormi, Viet-Hung~Nguyen, Alessandro~Kovtun, Samuel~Lara~Avila, Andrea~Liscio, Jean-Christophe~Charlier, Stephan~Roche, H\^{a}ldun~Sevin\c{c}li}

\begin{center}
	\fontsize{18}{30} \selectfont\sffamily{Supporting Information}\\
	\fontsize{20}{30} \selectfont\sffamily\textbf{\baslik}\\\vspace{10mm}
	\fontsize{14}{14} \selectfont\sffamily{\yazarlar}\\\vspace{5mm}
	\fontsize{12}{12} \selectfont\sffamily{E-mail: stephan.roche@icn2.cat; haldunsevincli@iyte.edu.tr}
\end{center}

\section{Kubo-Greenwood and Landauer-B\"uttiker methods}
An efficient linear scaling approach~\cite{fan_linear_2020} is used in Kubo transport to estimate the energy- and time-dependent mean squared displacement of the wave-packet that spreads into the investigated atomic structure,
\begin{equation}
\label{eqn:mean-square-displacement}
\Delta X^2(E,t) =  \frac{ \Trace{ \left[ \delta(E-\hat{H}) |\hat{X}(t) - \hat{X}(0)|^2 \right] } }{ g(E)},
\end{equation}
where $g(E)= \Trace [\delta(E - \hat{H})] $ is the density at energy $E$. The time-dependent semiclassical diffusion coefficient $D(E,t) = \frac{\partial }{\partial t} \Delta X^2(E,t)$ and its asymptotic limit $\tilde{D}(E)$ can then be calculated, allowing for the computation of both the electron conductivity and the mean free path as $\sigma(E) = e^2 g(E) \tilde{D}(E)$ and $\lmfp(E) = 2\tilde{D}(E)/v_F(E)$, 
respectively (with $v_F(E)$ being the carrier velocity).  
The conductivity values in rGO systems have been averaged over 10 different randomly chosen initial wave packets, and the calculation of the mean-square displacement was carried out through an efficient decomposition in terms of Chebyshev polynomials, with 5000 moments. Note that periodic boundary conditions are employed in both longitudinal and transverse directions.

Concerning the transport simulations performed using the Landauer-B\"uttiker technique, the rGO system is partitioned into three regions, namely the left and right electrodes (leads) and the central region. The {leads} are modeled as scattering free regions made up of the same {ideal} material. The Green function for the central region is calculated as $\GG(E)=\left[(E+i0^+)\II-\HH_C-\Sigma \right]^{-1}$, where $\II$ is identity matrix, $H_C$ is the Hamiltonian matrix for the central region, and the self-energy term includes effects of the left and right reservoirs as $\Sigma=\Sigma_L+\Sigma_R$. 
In this work, systems containing as many as $10^{6}$ atoms have been simulated, for which efficient decimation algorithms are implemented.~\cite{ryndyk_book_2015,sevincli_srep_2013}
The transmission amplitude is obtained from $\trans(E)=\Trace\left[\Gamma_L\GG\Gamma_R\GG^\dagger\right]$, where $\Gamma_{L(R)}=i [\Sigma_{L(R)}-\Sigma_{L(R)}^\dagger]$ are the left (right) broadening matrices.
Conductance values are calculated using the Landauer formula,
\begin{equation}
G=\frac{2e^2}{h}\int \left(- \frac{\partial f_{FD}(E,\mu,T)}{\partial E} \right) \mathcal{T}(E)  dE,
\end{equation}  
where $e$ is electron charge, $f_{FD}$ is Fermi-Dirac distribution function, $T$ is temperature (herewith 100~K) and $\mathcal{T}(E)$ is the transmission probability for a given energy $E$. 
Since low-energy properties are only of interest, transmission coefficients are integrated over 40 $k$-points in the transverse direction in order to reach sufficiently accurate energy resolution. At last, a geometric average on the transmission function over an ensemble of 20 samples is applied to overcome the sample size effects.

\clearpage
\section{Supplementary details on simulated structures}
\haldun{
The types and amounts of defects are deduced from MD simulations replicating the thermal annealing of GO on a computer. In particular, following the protocol in Ref. 6, several atomistic samples of GO have been generated with a total number of atoms as large as 10000 and an initial oxygen concentration of 35\%. The thermal reduction of the systems has been simulated for different annealing temperatures and a statistical analysis of the chemical and morphological properties of the resulting rGO samples has been performed. In the table below, a summary of the values of the most relevant chemical species observed in the the samples after reduction is given. At the annealing temperature of 900oC, a final concentration of oxidizing agents of 5\% has been found, in extremely good agreement with the experimental samples.  From a detailed exploration of the atomistic structures, the concentration and type of defects has been derived and used to model rGO samples for transport calculations.
}

\begin{table}[]
	\haldun{
	\begin{tabular}{lccccccc}
		\hline
		& C sp²  & C sp³  & C-OH  & C-O-C  & C=O  & O-C=O & O/CFIT         \\
		& (\%) &  (\%) & (\%) &  (\%) &  (\%) &  (\%) &          \\
		\hline
		GO         & 32.9       & 11.5       & 3.0       & 43         & 6.5      & 3.0        & 0.35 (imposed) \\
		rGO 300 °C & 75.4       & 10.5       & 2.9       & 3.9        & 4.8      & 2.5        & 0.154          \\
		rGO 600 °C & 83.5       & 8.8        & 1.1       & 3.7        & 1.1      & 1.8        & 0.11           \\
		rGO 900 °C & 94.1       & 3.6        & 0.3       & 0.8        & 0.6      & 0.6        & 0.05          \\
		\hline		
	\end{tabular}
	}
	\caption{
		\haldun{Concentration of different chemical species observed in the atomistic samples of rGO after the reduction process at different annealing temperatures.
			}
	}
\end{table}

\clearpage

\section{Supplementary details regarding tight-binding parameters}

As explained in the main text, while the change in C-C bond length is modeled by the distance dependence of hopping energies, the local doping due to the localized states induced by defects and impurities are included by adding on-site energies to C-atoms  surrounding their position. These on-site energies have the following common form
\begin{itemize}
	\item $\varepsilon_D$ applied to C atoms directly connected to impurities/defects
	\item $\varepsilon_n$ applied to other surrounding C atoms
\end{itemize}
\begin{eqnarray}
	\varepsilon_n =\frac{\varepsilon_0}{1+(d_n/\lambda_D)^\kappa}
\end{eqnarray}
where $d_n$ is the distance from the $n^{th}$ C atom to the considered impurity, $\varepsilon_0$ is the maximum value of $\varepsilon_n$,     $\lambda_D$ is the decay length and the number $\kappa$ is determined depending on the defect/impurity types.

\vspace{10mm}
\noindent
\begin{tabular}{lllllc}
\hline
Oxygen impurity	& $\lambda_D=1.0$~$\textup\AA$ & $\kappa=3$ &$\varepsilon_{0}=1.6$~eV &$\varepsilon_{D}=-28.0$~eV  &    \\ 
OH group		 & $\lambda_D=1.0$~$\textup\AA$& $\kappa=3$ &$\varepsilon_{0}=1.8$~eV &$\varepsilon_{D}=-28.0$~eV  &    \\ 
585 defect	    & $\lambda_D=5.0$~$\textup\AA$ & $\kappa=5$ &$\varepsilon_{0}=1.0$~eV  &$\varepsilon_{D}=-1.3$~eV   &    \\  
Stone-Waled defect	& $\lambda_D=1.6$~$\textup\AA$ & $\kappa=5$ &$\varepsilon_{0}=-1.5$~eV  &$\varepsilon_{D}=-1.2$~eV  &    \\ 
555-777 defect &	$\lambda_D=8.0$~$\textup\AA$	& $\kappa=5$& $\varepsilon_{0}=0.4$~eV& $\varepsilon_{D1}=-1.8$~eV  & $\varepsilon_{D2}=-0.5$~eV     \\
\hline
\end{tabular} 
\vspace{5mm}

The electronic band structures obtained using the proposed tight binding Hamiltonians are presented in Figs.~\ref{fig_supp2} and \ref{fig_supp3}. Indeed, our proposed tight-binding models reproduce well the low energy bands, compared to the DFT results. 

\begin{figure}
	\includegraphics[width=1\textwidth]{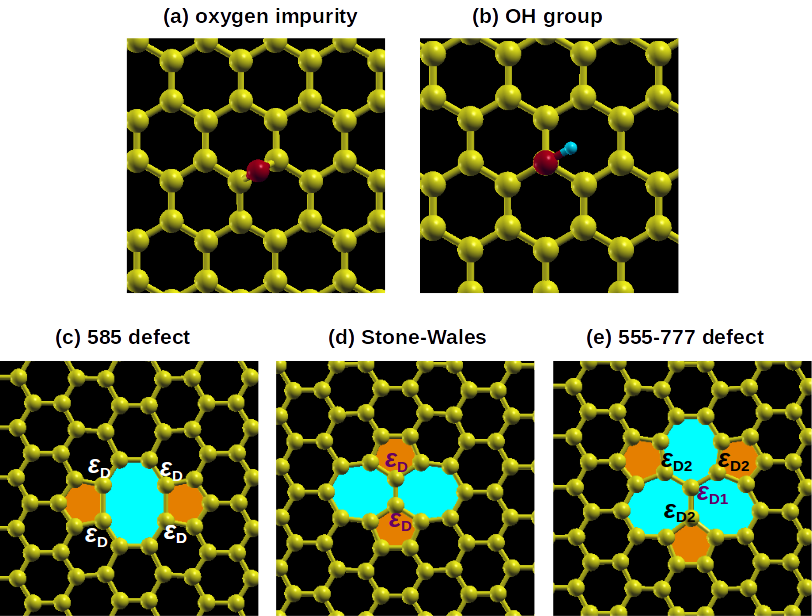}
	\caption{Defects and impurities investigated in this work.}
	\label{fig_supp1}
\end{figure}

\begin{figure}
	\includegraphics[width=1.\textwidth]{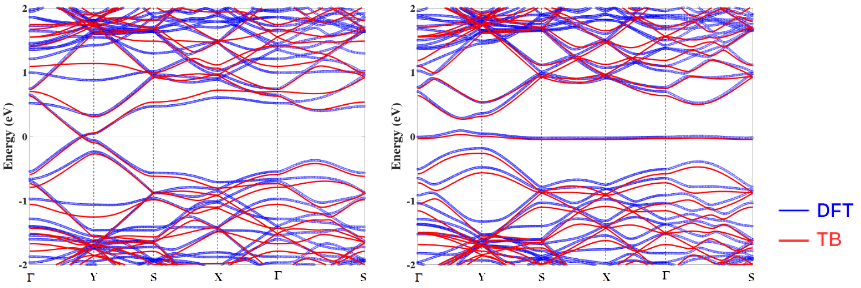}
	\caption{The electronic band structure of graphene with oxygen impurity (left) and OH group (right). TB calculations fit to the DFT results.}
	\label{fig_supp2}
\end{figure}

\begin{figure}
	\includegraphics[width=1.\textwidth]{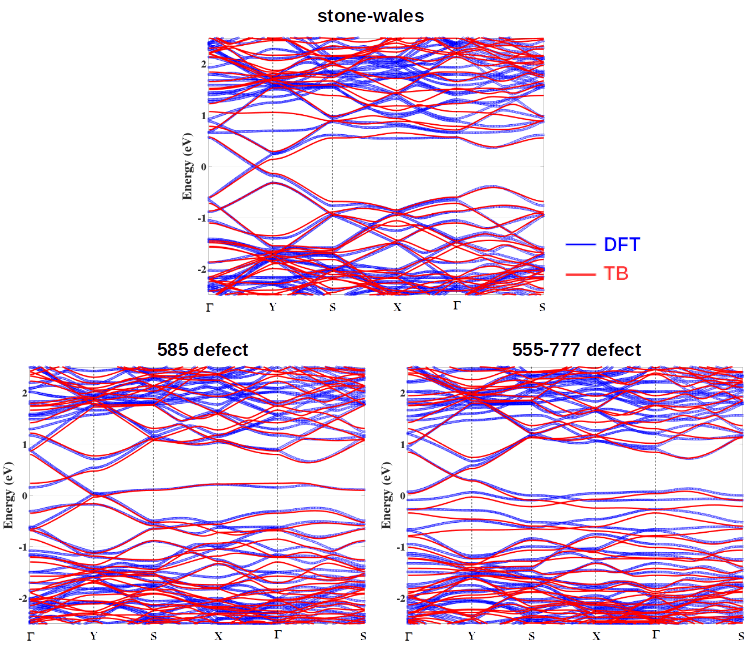}
	\caption{The electronic bandstructure of graphene with structural defects.
		TB calculations fit to the DFT results.}
	\label{fig_supp3}
\end{figure}

\clearpage

\begin{figure}
	\includegraphics[width=.7\textwidth]{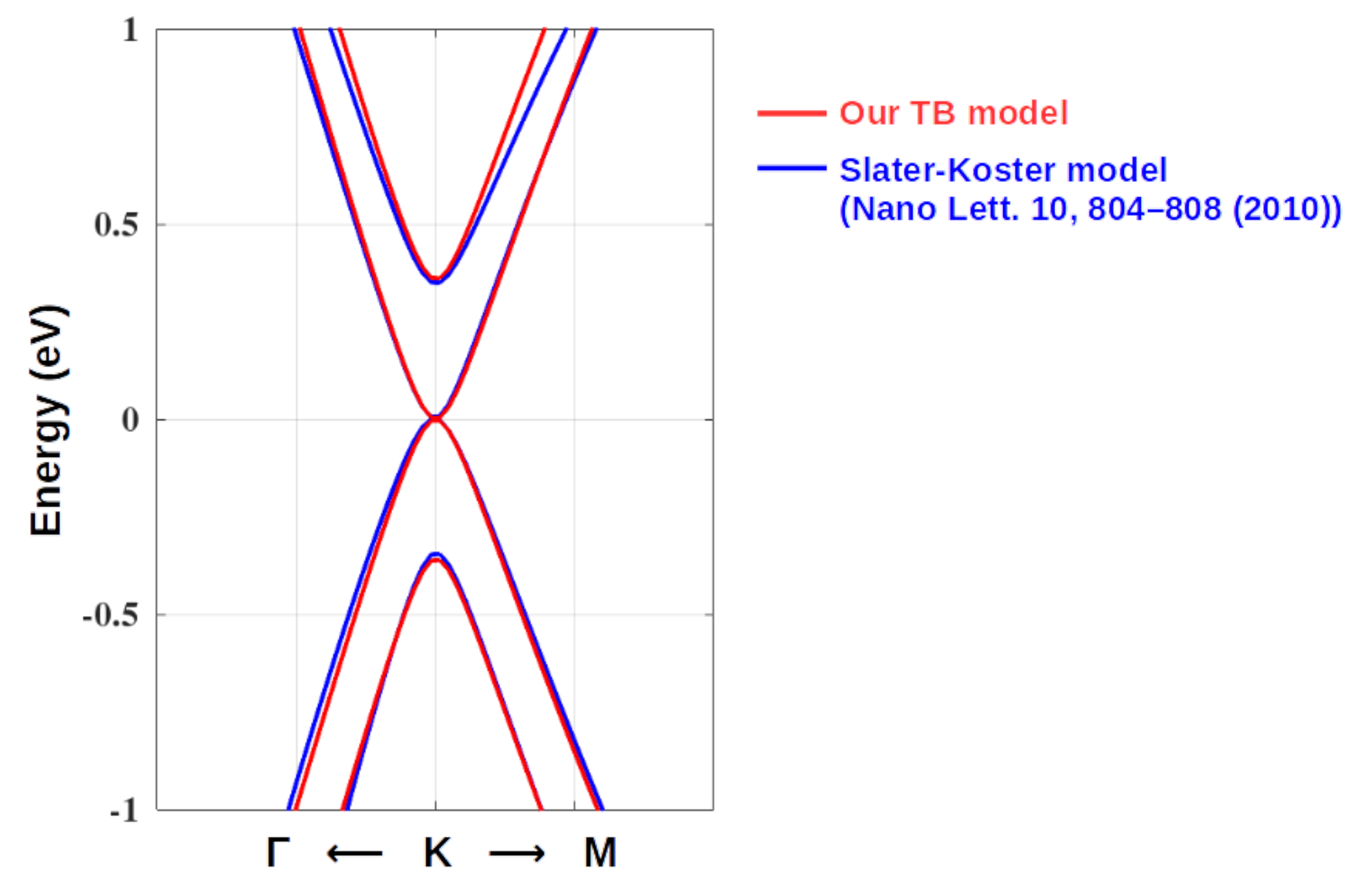}
	\caption{
		\haldun{
		Comparison of the electronic bands of bilayer graphene obtained using our TB Hamiltonian and Slater-Koster like models in Nano Lett. 10, 804–808 (2010).}
	}
	\label{fig:slater-koster}
\end{figure}

\haldun{
Other sophisticated models (i.e., larger distance neighbor as well as Slater-Koster like models) generally present a disadvantage that a large number of adjusted parameters are required to model accurately the considered defective systems. This disadvantage also gives rise to some difficulties for the implementation of transport calculations in the large scale devices while the accuracy is not significantly improved.
In Fig.~\ref{fig:slater-koster} a comparison of the computed electronic band strucures of bilayer graphene obtained using our used model and Slater-Koster like models in Ref.~\citenum{laissardiere:nanolett_2010} that has been shown to compute well the electronic structure of both Bernal stacking and twisted bilayer graphene systems.}

\clearpage
\section{Note on parametrization and system setups}

The tight-binding parametrization for structural defects involves modifying onsite energies of $sp^2$ orbitals by a value which decays exponentially with the distance, and onsite energies due to defects surrounding a particular atom are taken additive. For Landuer-B\"uttiker calculations, we also introduced defect-free buffers of 2.57~nm in length between the scattering region and the leads to saturate the effects due to the onsite energies (see Fig.\ref{fig1} right panel), and cutoff radius of 2.57~nm was used as the range of the modification. For the scaling analysis, the central transport channel is lengthened by adding 1.284~nm-length blocks (which corresponds to a mesh resolution of the same length). 

\clearpage
\section{Supplementary mean-free-path plots}

\begin{figure}[h]
	\includegraphics[width=1\textwidth]{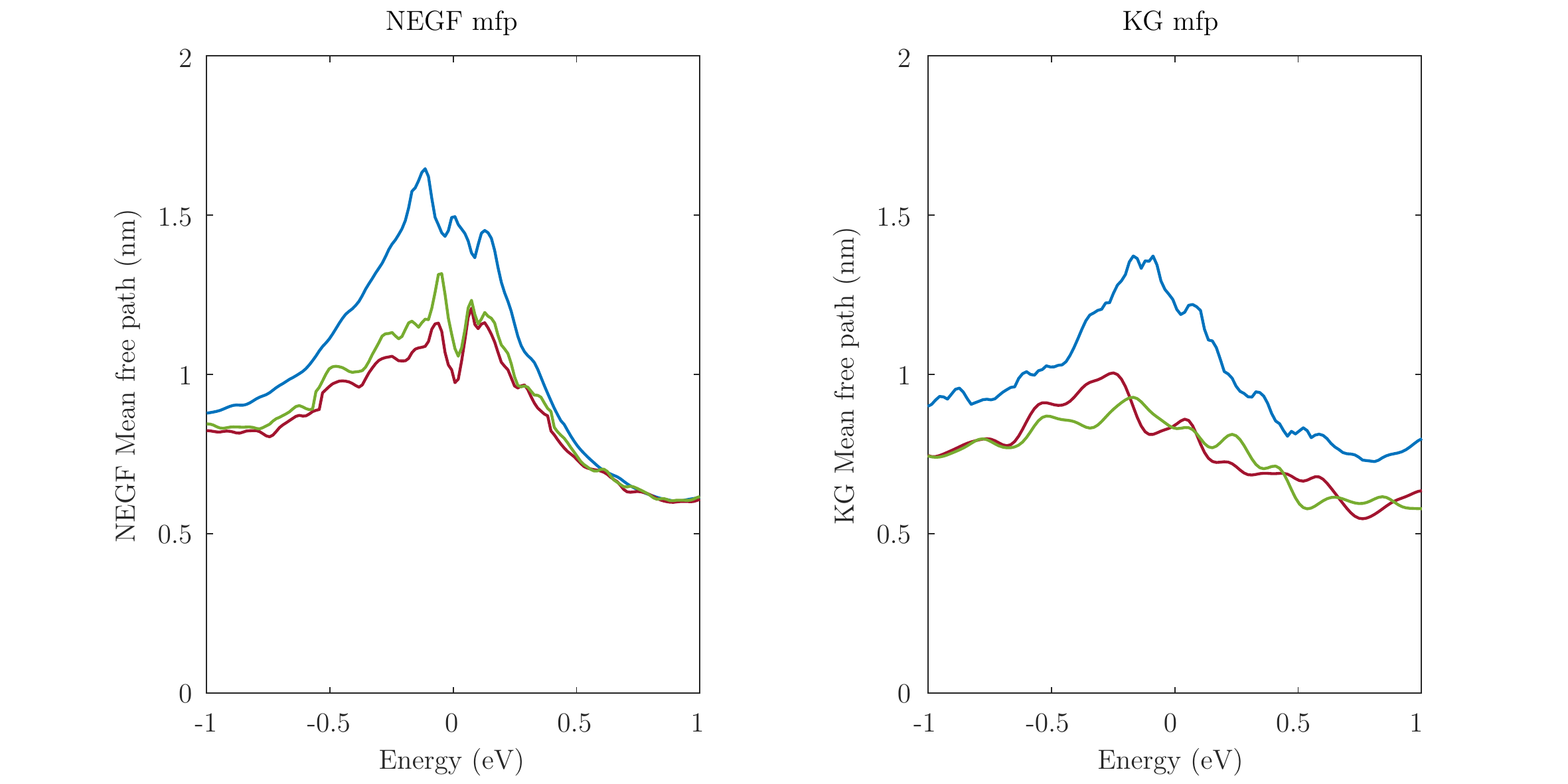}
	\caption{
		Mean-free-paths for different numbers of layers as obtained from NEGF and KG simulations are shown. Blue, red and green curves represent mono/bi/tri-layer rGO systems.
	}
	\label{fig_supp_mfp}
\end{figure}

Mean-free-paths as obtained from Landauer-B\"uttiker and Kubo-Greenwood simulations with changing layer thickness are shown for comparison in Figure~\ref{fig_supp_mfp}.
The agreement between two methodologies is remarkable.

\clearpage
\section{Supplementary results with lower defect concentrations}

\begin{figure}[h]
	\includegraphics[width=.75\textwidth]{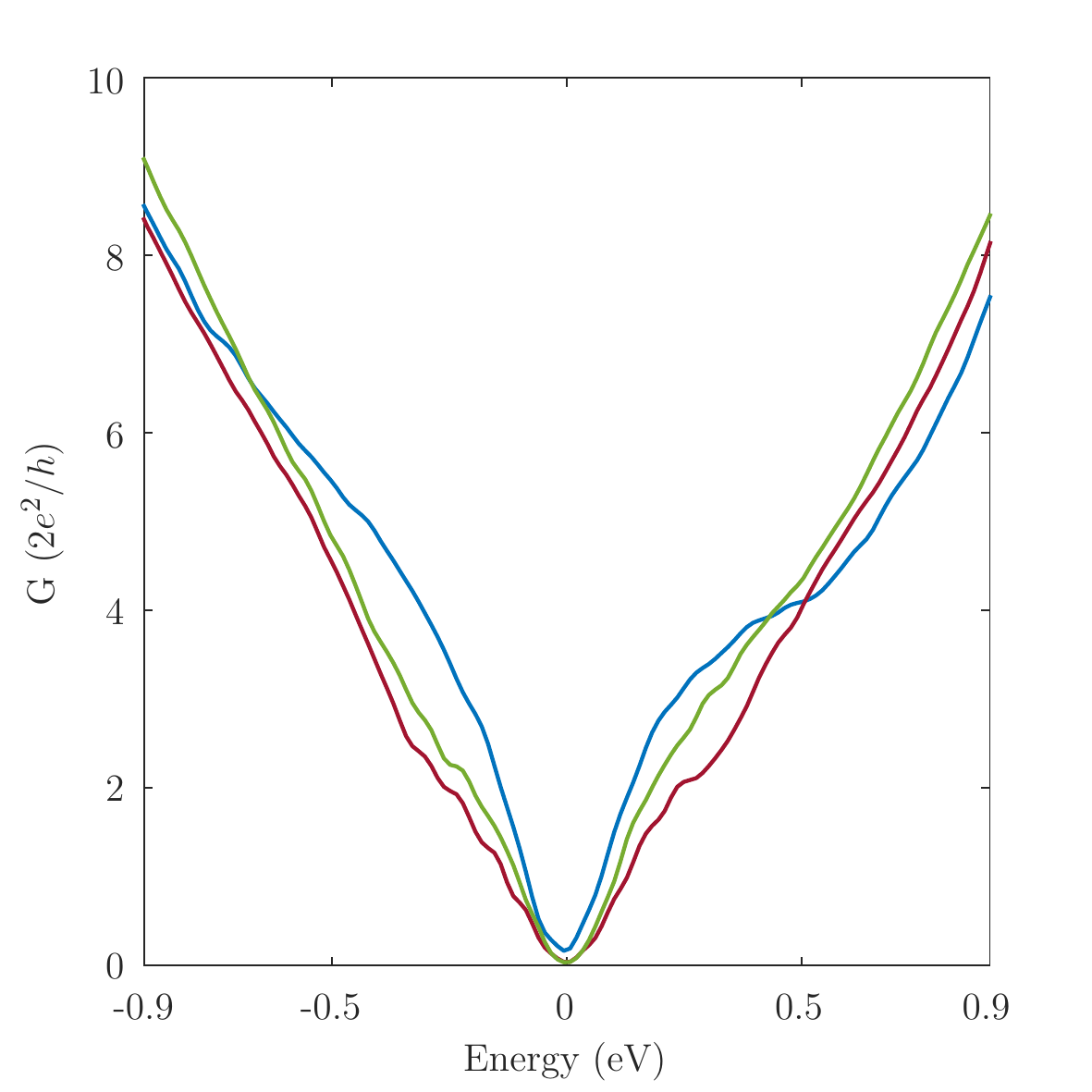}
	\caption{Transmission spectrum through rGO with low concentration of imperfections. The device sizes are  the same with those in the main text, impurity concentration is 10 times lower with 99.5\% sp$^2$ ratio. Blue, red and green curves correspond to mono/bi/trilayer rGO, respectively.}
	\label{fig_supp_lowdefect}
\end{figure}

We have shown in the main text that interlayer coupling affects charge transport of defect-free and defective systems in opposite ways. Namely, in defect-free systems conductance is reduced with the number of layers, whereas it is enhanced in rGO, which contains 95\% sp$^2$ carbon. Reducing the amount of impurities, it is possible to observe the transition. In Fig.~\ref{fig_supp_lowdefect}, defect concentration is 10 times lower than those in the main text (99.5\% sp$^2$ carbon), where monolayer is observed to have the highest transmission values around the charge neutrality point.

\clearpage
\section{Supplementary LDOS plots}

\begin{figure}[h]
	\includegraphics[width=.5\textwidth]{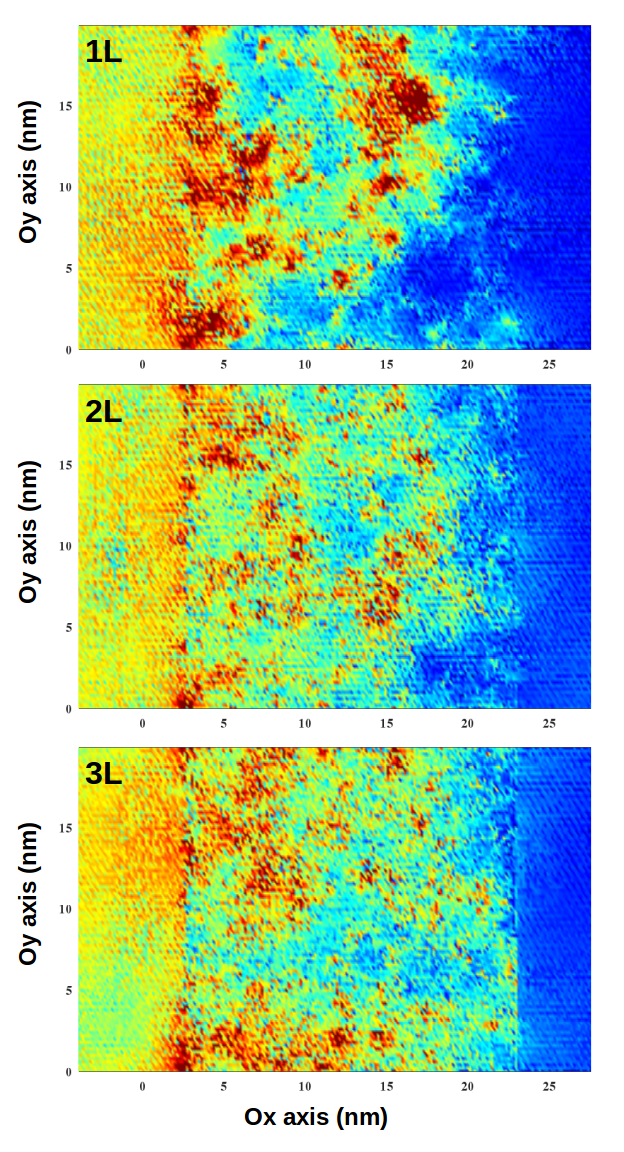}
	\caption{
		Left injected LDOS plots for mono/bi/tri-layer rGO. High and low LODS values are distinguihed on the  left/right electrodes. Localization is more pronounced in monolayer, whereas the carrier density more dispersed in trilayer sample.
		}
	\label{fig_supp_ldos}
\end{figure}
Using the Green's function method, we could compute the left (right, respectively) injected LDOS \cite{datta_nanoscale_2000}, reflecting the propagation of electrons from left to right (right to left, respectively) electrodes. In particular, the left-injected LDOS is given by
\begin{eqnarray}
	LDOS_L = \frac{G\Gamma_LG^\dagger}{2\pi}.
\end{eqnarray}
LDOS in the mulitlayer zones are averaged over the layers to show the contributions from all layers. The decay of the presented left-injected LDOS along the Ox axis in Fig.S6 is essentially due to scatterings with defects/impurities, manifesting as the electronic localization in the device region. In the monolayer case, charge localization is more pronounced than bilayer and trilayer systems, in agreement with Fig.~\ref{fig2}f. Importantly, it is shown that the improved propagation of electrons from left to right electrodes is obtained when increasing number of graphene layers, thus illustrating the transport properties discussed in the main text.

\clearpage
\section{Temperature-dependence of electrical resistivity $\rho(T)$, Efros-Shklovskii variable range hopping model}

At low temperature, charge transport in graphene-based materials is typically occurring via charge hopping in a disorder-broadened density of states near the Fermi level $g(E_F)$.~\cite{cheah_jpcm_2013} In the Ohmic regime, the resistivity  is tipycally modelled by a stretched exponential behavior:
\begin{eqnarray}
    \label{eqn:vrh}
    \rho(T)=\rho_{0,VRH} \exp
    \left(
    \frac{T_0}{T}
    \right)^\beta,
\end{eqnarray}
where $\rho_{0,VRH}$ is a prefactor and $\beta$ is a characteristic exponent. $T_0$ represents a characteristic temperature correlated to the localization length $(\lloc)$, the higher the first one, the lower the latter. The $\lloc$ is defined as the average spatial extension of the charge carrier wave function: the lower the $T_0$, the larger the $\lloc$. 

\begin{figure}
	\includegraphics[width=0.99\textwidth]{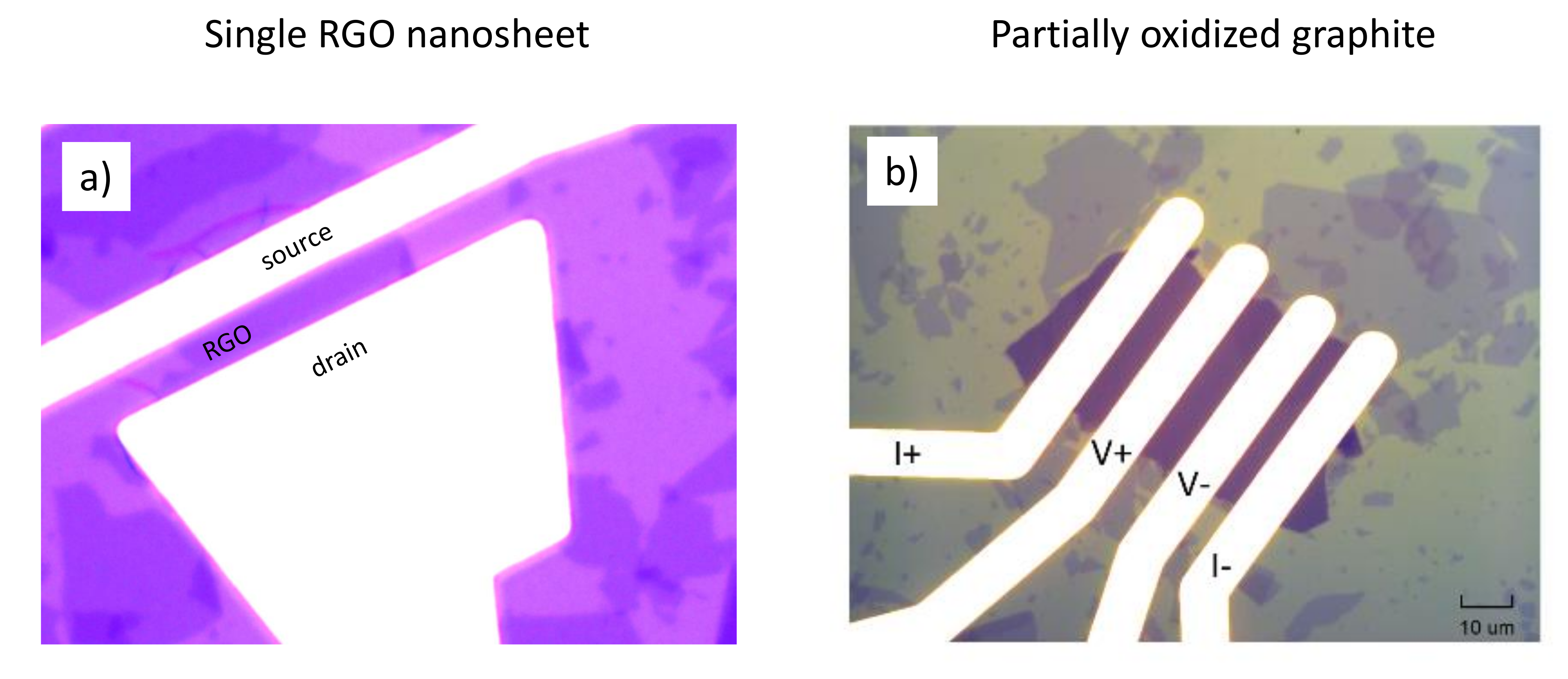}
	\caption{
		\ciz{
		Sample thickness was measured by atomic force microscopy (AFM) and the graphitic structure is confirmed by Raman measurements.}
		\haldun{
		Optical images of devices fabricated using (a) single rGO nanosheet and (b) partially oxidized graphite. Samples thickness were measured by atomic force microscopy (AFM) and the graphitic structure of (b) was confirmed by Raman measurements.}
	}
	\label{fig:supp_oxidized-graphite}
\end{figure}

\begin{figure}[t]
	\centering
	\includegraphics[width=1.0\textwidth]{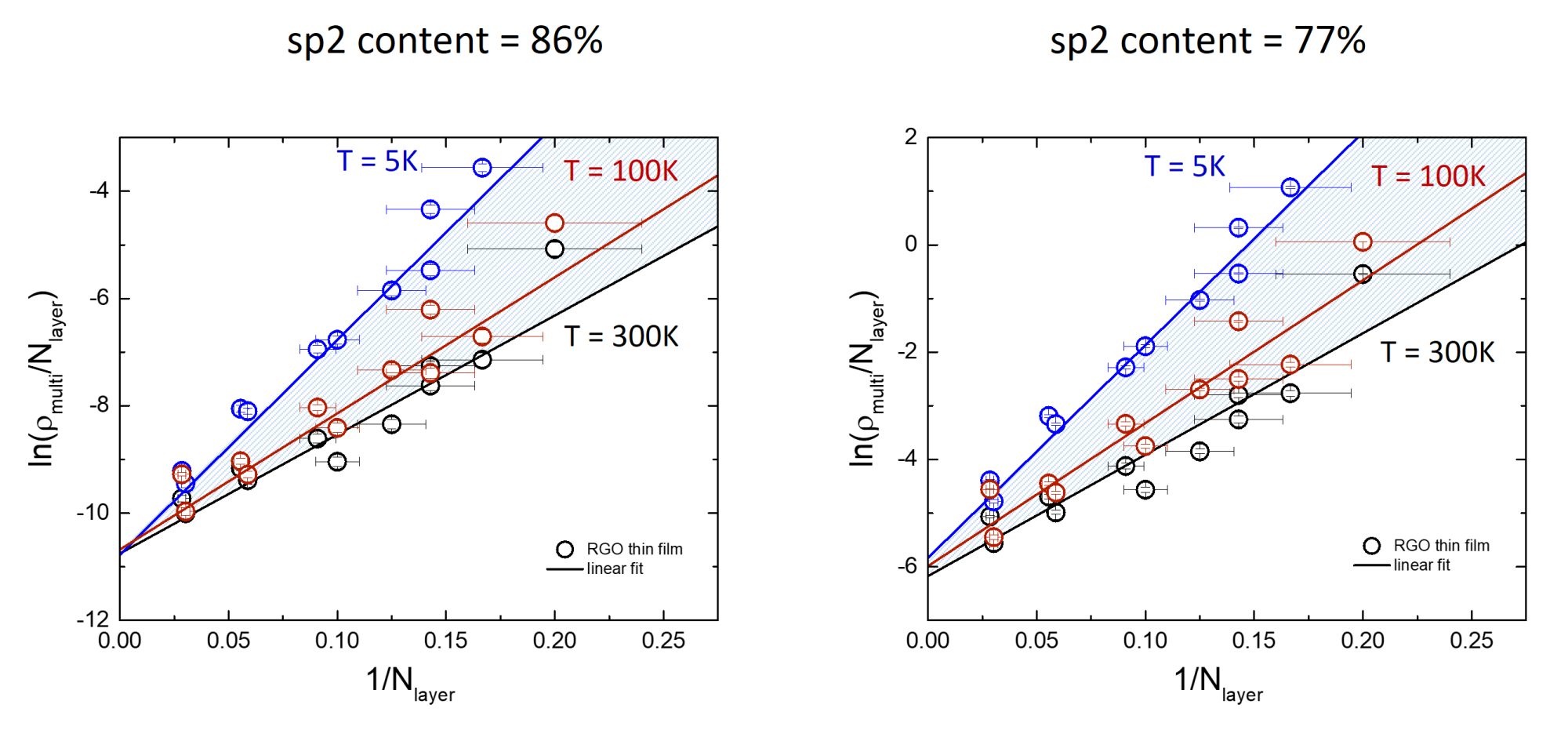}
	\caption{Correlation plots. Experimental data of multilayer RGO (circles for devices reported in Ref.~\citenum{kovtun_acsnano_2021} and acquired at different temperatures. Samples with both sp$^2$ contents (86\% and 77\%) show linear dependence, in good agreement with the theoretical prediction for scaling (cf. Eqn.~\ref{eqn:multilayerscaling} in the main text). All the linear fitting curves calculated at different temperatures are included between the two curves acquired at 5~K and 300~K (dashed area).}
	\label{fig:supp_correlation}
\end{figure}

The analytic expression reported in Eqn.~\ref{eqn:vrh} is quite general depending on the model commonly referred as variable range hopping (VRH). The stretching exponent $\beta$ is strongly dependent on the shape of $g(E_F)$, e.g. when the density of states is constant (Mott-VRH model),~\cite{mott_cjp_1956} the $\beta$ value directly depends on the system’s dimension ($D$) with the form $\beta= 1/(D+1)$. Reduced graphene oxide thin films show the presence of a gap at the Fermi level due to the Coulomb interaction between the occupied, excited state above $E_F$ and the hole left by the same electron below $E_F$. This case is described by the so called Efros Shklovskii model (ES-VRH)~\cite{efros_jpcm_1975}
with characteristic exponent of Eqn.~\ref{eqn:vrh} $\beta=1/2$, which does not depend on the system dimensionality. The characteristic temperature $T_0$ for 2D materials is given by
\begin{eqnarray}
    \label{eqn:T0}
    T_0=\frac{2.8e^2}{4\pi\epsilon_0\epsilon_rk_B\lloc}
    =\frac{1}{A\epsilon_r\lloc},
\end{eqnarray}
where $e$ is the elementary charge, $\epsilon_0$ and $\epsilon_r$ represent the vacuum permittivity and the relative permittivity of the material and $k_B$ is the Boltzmann constant. For the sake of simplicity, all the universal constants are collected by the parameter $A = 0.021 \mu\mathrm{m}^{-1}\mathrm{K}^{-1}$.
Combining Eqns.~\ref{eqn:vrh} and \ref{eqn:T0}, we obtain the mathematical expression reported in the main text,
\begin{eqnarray}
        \rho(T)=\rho_0\exp\left(\frac{1}{A\epsilon_r\lloc T}\right)^{1/2}.
\end{eqnarray}

\clearpage
\section{Electrical resistivity measurements $\rho(T)$}

\haldun{
	Single rGO nanosheet and partially oxidized graphite were prepared by thermal annealing ($T_\mathrm{ann} = 900^o$C) of GO and oxidized nanographite, respectively, deposited on clean SiO$_2$/Si substrates (2,000 rpm for 60s).
}

\haldun{
	The micrometric electrodes were lithographically patterned to characterize the electrical transport across a limited number of overlapping flakes. Lithography was carried out by exposing a standard photoresist (AZ1505, Microchemicals) with the 405~nm laser of a Microtech laser writer. A 30-nm-thick Au film (without adhesion layer) was thermally evaporated onto the patterned photoresist and lift-off was carried out in warm acetone (40$^o$C).
}

\haldun{
	The resistance vs temperature measurements were carried out with a Quantum Design Physical Properties Measurements System (PPMS), using an external Keithley 2636 Source-Meter. The resistance was measured in the temperature range between 300~K to 5~K with a slow ramp (1~K/min). The Ohmic behavior of the device was checked by the linearity of the I-V curves.
}

\begin{figure}[t]
	\includegraphics[width=14cm]{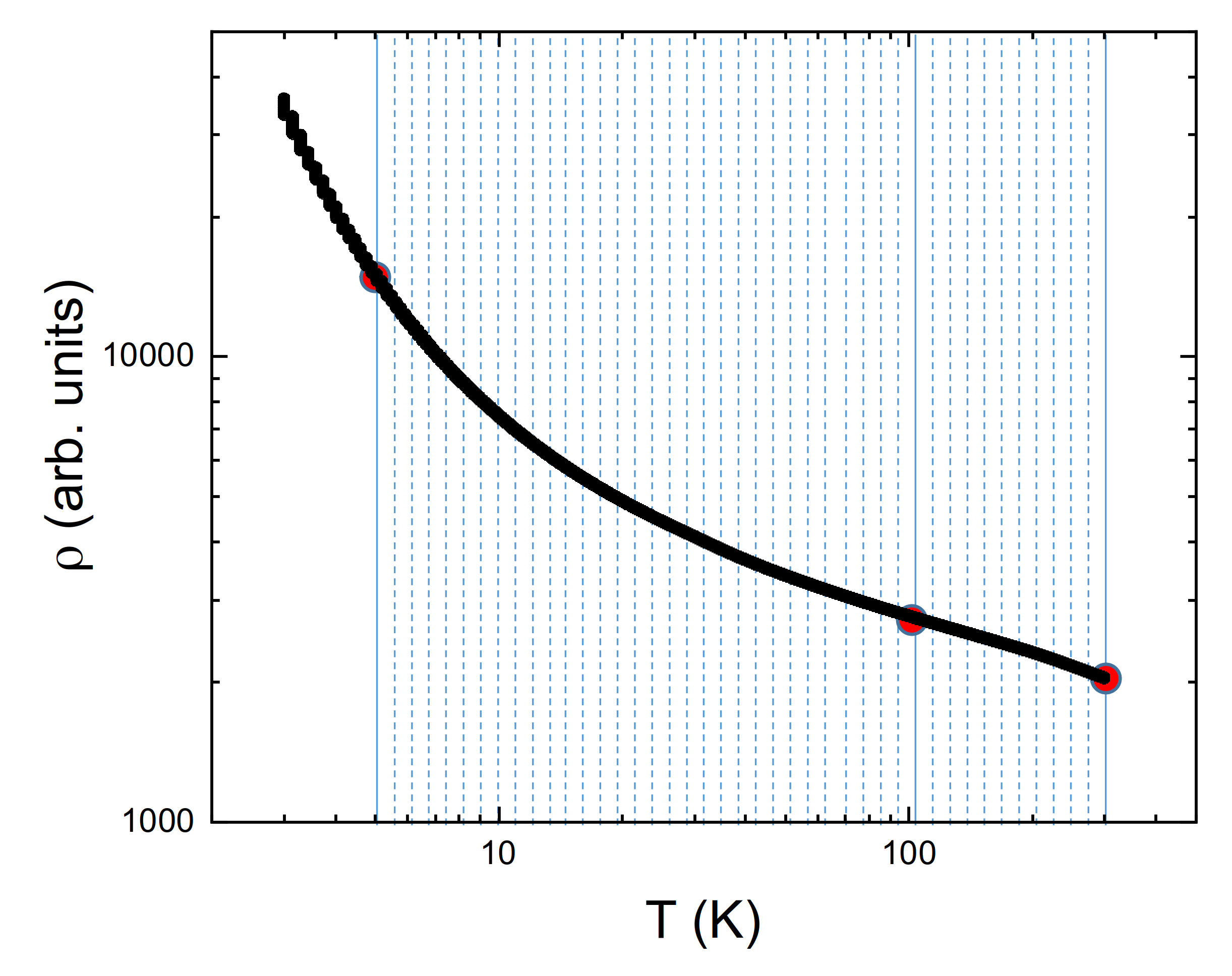}
	\caption{
		\haldun{Example of measured $\rho(T)$ curve and sampled temperature values. Red circles correspond to the resistivity values depicted in the correlation plots.}
	}
	\label{fig:supp_rho-logT}
\end{figure}

\haldun{
	Typically, each acquired $\rho(T)$ curve corresponds to an array of $>$50,000 resistivity values at different temperatures. For sake of simplicity, such $\rho(T)$ curves were sampled at 43~temperature values with logarithmic steps, as reported in Fig.~\ref{fig:supp_rho-logT}. Three values acquired at 5~K, 100~K and 300~K are reported in Figure~5 in the main text. 
}

\begin{landscape}
	\begin{table}[]
		\haldun{
			\caption{Summary of experimental parameters of all the studied devices}
			\begin{tabular}{|c|l|l|l|c|c|c|}
				\hline sample &  & $N_\textrm{layer}$ & $\xi$ (nm) &  $\rho_\textrm{5K}$ & $\rho_\textrm{100K}$ & $\rho_\textrm{300K}$ \\
				\hline \multirow{3}{8em} {single RGO nanosheet}
				& \multirow{3}{5em}{Device 28 [ref.5 main text]} & 1 & 4.0$\pm$0.3 &  &  & \\
				&  &  &  &  &  & \\
				&  &  &  &  &  & \\
				&  & 1 & 3.7$\pm$0.4 &  &  & \\
				\hline \multirow{2}{8em}{Bi-layer RGO}
				& Device 26 & 2 & 7.5$\pm$0.8 &  &  & \\
				& Device 27 & 2 & 7.8$\pm$1.4 &  &  & \\
				\hline \multirow{2}{8em}{RGO thin film}
				& Device 6 & 5$\pm$1 & 18$\pm$2 & -- & $(3.6\pm 12)\times 10^{-4}$ & $(2.8\pm0.9)\times 10^{-4}$ \\
				& Device 1 & $6\pm1$  & $26\pm4$ & $(13\pm3)\times 10^{-4}$ & $(6.5\pm1.7)\times10^{-5}$ & $(4.7\pm1.3)\times10^{-4}$ \\
				\hline \multirow{2}{8em}{partially oxidized graphite}
				&  & \multirow{2}{*}{$6\pm1$} & \multirow{2}{*}{$243\pm56$} & \multirow{2}{*}{$(5.9\pm1.3)\times10^{-4}$} & \multirow{2}{*}{$1.1\pm0.3)\times10^{-4}$} & \multirow{2}{*}{$(2.9\pm0.7)\times10^{-5}$} \\
				&  &  &  &  &  & \\
				\hline \multirow{9}{8em}{RGO thin film}
				& Device 7  & $7\pm1$  &  & $(3.1\pm0.7)\times10^{-4}$ & $(3.4\pm0.8)\times10^{-5}$ & $(2.4\pm0.6)\times10^{-5}$ \\
				& Device 17 & $7\pm1$  &  & $(6.4\pm1.4)\times10^{-4}$ & $(1.1\pm0.2)\times10^{-4}$ & $(4.8\pm1.1)\times10^{-4}$ \\
				& Device 2  & $8\pm1$  &  & $(2.4\pm0.5)\times10^{-4}$ & $(3.7\pm0.7)\times10^{-5}$ & $(1.9\pm0.4)\times10^{-5}$ \\
				& Device 8  & $10\pm1$ &  & $(9.7\pm1.5)\times10^{-5}$ & $(1.6\pm0.2)\times10^{-5}$ & $(1.2\pm0.2)\times10^{-5}$ \\
				& Device 3  & $11\pm1$ &  & $(7.9\pm1.2)\times10^{-5}$ & $(2.9\pm0.4)\times10^{-5}$ & $(1.5\pm0.2)\times10^{-5}$ \\
				& Device 9  & $17\pm1$ &  & $(3.7\pm0.4)\times10^{-5}$ & $(1.3\pm0.1)\times10^{-5}$ & $(1.3\pm0.1)\times10^{-5}$ \\
				& Device 4  & $18\pm1$ &  & $(4.3\pm0.4)\times10^{-5}$ & $(1.6\pm0.2)\times10^{-5}$ & $(1.4\pm0.1)\times10^{-5}$ \\
				& Device 10 & $33\pm2$ &  & $(2.8\pm0.3)\times10^{-5}$ & $(1.2\pm0.1)\times10^{-5}$ & $(1.3\pm0.1)\times10^{-5}$ \\
				& Device 5  & $35\pm2$ &  & $(2.4\pm0.2)\times10^{-5}$ & $(2.0\pm0.2)\times10^{-5}$ & $(1.4\pm0.2)\times10^{-5}$ \\
				\hline
			\end{tabular}
		}
	\end{table}
\end{landscape}

\end{document}